# The Cassiopeia A Supernova was of Type IIb

Oliver Krause[1*], Stephan M. Birkmann[1], Tomonori Usuda[2], Takashi Hattori[2], Miwa Goto[1], George H. Rieke[3] & Karl A. Misselt[3]

[1]*Max-Planck-Institut für Astronomie, Königstuhl 17, 69117 Heidelberg, Germany*

[2]*National Astronomical Observatory of Japan, 650 North A'ohoku Place, Hilo, Hawaii, USA*

[3]*Steward Observatory, 933 N Cherry Ave, Tucson, Arizona, USA*

*To whom correspondence should be addressed: Email: krause@mpia.de

**Cassiopeia A is the youngest supernova remnant known in the Milky Way and a unique laboratory for supernova physics. We present an optical spectrum of the Cassiopeia A supernova near maximum brightness, obtained from observations of a scattered light echo – more than three centuries after the direct light of the explosion swept past Earth. The spectrum shows that Cassiopeia A was a type IIb supernova and originated from the collapse of the helium core of a red supergiant that had lost most of its hydrogen envelope prior to exploding. Our finding concludes a long-standing debate on the Cassiopeia A progenitor and provides new insight into supernova physics by linking the properties of the explosion to the wealth of knowledge about its remnant.**

The supernova remnant Cassiopeia A is one of the most studied objects in the sky with observations from the longest radio waves to gamma-rays. From the remnant expansion rate it is estimated that the core of its progenitor star collapsed around the year 1681 +/- 19, as viewed from Earth (*1*). Due to its youth and proximity of 3.4 $^{+0.3}_{-0.1}$ kpc (*2*), Cas A provides a unique opportunity to probe the death of a massive star and to test theoretical models of core-collapse supernovae. However, such tests are compromised because the Cas A supernova showed at most a faint optical display on Earth at the time of explosion. The lack of a definitive sighting means there is virtually no direct information about the type of the explosion and the true nature of its progenitor star has been a puzzle since the discovery of the remnant (*3*).

The discovery of light echoes due both to scattering and to absorption and re-emission of the outgoing supernova flash (*4, 5*) by the interstellar dust near the remnant raised the possibility of conducting a post-mortem study of the last historic Galactic supernova by observing its scattered light. Similarly, the determination of a supernova spectral type long after its explosion utilizing light echoes has recently been demonstrated for an extragalactic supernova (*6*).

We have monitored infrared echoes around Cas A at a wavelength of 24 μm using the MIPS instrument aboard the Spitzer Space Telescope (*4*). The results confirm that they arise from the flash emitted in the initial explosion of Cas A (*5*). An image taken on 20 August 2007 revealed a bright (flux density $F_{24\mu m}$ = 0.36 +/- 0.04 Jy ; 1 Jy ≡ $10^{-26}$ W $m^{-2}$ $Hz^{-1}$) and mainly unresolved echo feature located 80 arcmin north-west of Cas A (position angle 311° east of north). It had not been detected ( $F_{24\mu m}$ < 2 mJy; 5- ) on two previous images of this region obtained on 2 October 2006 and 23 January 2007 (Fig. 1).

An image obtained on 7 January 2008 shows that the peak of the echo has dropped in surface brightness by a factor of 18 and shifted towards the west. Transient optical emission associated with the infrared echo was detected in an R-band image obtained at a wavelength of 6500 Å at the Calar Alto 2.2m telescope on October 6, 2007 with a peak surface brightness R = 23.4 +/- 0.2 mag arcsec$^{-2}$. No optical emission feature down to surface brightness limit R = 25.1 mag arcsec$^{-2}$ (3 ) was detected towards this position in a previous R-band image obtained at the Steward Observatory 90-inch telescope on 18 September 2006.

We have acquired a deep R-band image of the echo with the FOCAS instrument at the SUBARU telescope on 9 October 2007 (Fig. 2). The morphology of the optical emission in this image closely matches the mid-infrared one observed 50 days earlier, with the echo resolved into compact emission knots of about 2 arcsec diameter in the R-band image. A long-slit spectrum covering the northern one of these compact knots (Fig. 2B) was obtained with FOCAS on the same night, covering the wavelength range from 4,760 to 9,890 Å with a spectral resolution of 24 Å.

This echo spectrum unambiguously shows light of a supernova origin (Fig. 3): broad emission lines with P-Cygni absorption components from neutral and singly ionized elements are detected, all of which are commonly observed in core-collapse supernovae. A prominent feature of the Cas A supernova spectrum is an H emission line with a width of 17,000 km/s at half maximum and a blue-shifted absorption minimum at -11,000 km/s. Other strong lines are Na I D, the Ca II infrared triplet and permitted emission lines of neutral oxygen at 7774 and 9264 Å. In addition, lines of He I 7065 Å and likely also He I 5876 Å (blended with Na I D) and He I 7821 Å (blended with [Ca II] 7291, 7324 Å) are detected.

While the presence of the hydrogen line classifies Cas A as type II supernova, the additional appearance of weak helium lines is characteristic of the rare class of type IIb supernovae (SNe). They originate from the core-collapse of massive stars that have lost most of their hydrogen envelopes prior to exploding and consist of a nearly bare helium core at the time of collapse. SNe IIb initially show a type II spectrum dominated by their hydrogen-poor envelope and gradually transform into a SN Ib spectrum from the inner helium-core (*7, 8, 9*). A type IIb for the Cas A supernova is supported by comparing its spectrum with that of the prototypical type SN IIb 1993J (*10*) (Fig. 3), the collapse of a red supergiant[11] with a main sequence mass of 13-20 M$_\odot$ (*12, 13*) in the nearby galaxy M81. The spectra are remarkably similar in terms of the presence of important spectral features and their strengths. Since the echo spectrum visible on Earth represents supernova light over an interval of time around maximum brightness due to the spatial extent of the interstellar cloud, the comparison spectrum was derived as the time-average of brightness-weighted spectra of SN 1993J obtained during days 1 to 83 after collapse (*10, 14*), consistent with the geometrical constraints of the echo (Fig. S1).

The identity of the Cas A progenitor has been the subject of tremendous debate (*15*). A type II/IIb has been suggested for the Cas A supernova (*16, 17*), based on the detection of a few remnant ejecta knots containing some hydrogen at space velocities between 9,000 and 10,300 km/s (*17, 18*). The flat-top shape of H emission in our Cas A spectrum is consistent with a thin hydrogen-rich shell above the photosphere and expanding at about 10,000 km/s. In contrast, the overall lack of hydrogen emission in most knots and the nitrogen enrichment in the remnant were widely interpreted as signatures of the collapse of a Wolf-Rayet star in a type Ib supernova (*18*). The SN Ib 2005bf was likely produced by a nitrogen-rich Wolf-Rayet star and has been proposed for a possible template for the Cas A supernova (*19*). However, the spectrum of SN 2005bf does not provide a good match to our spectrum of the Cas A supernova.

Further evidence for a red supergiant progenitor comes from the comparison of SN 1993J with Cas A (*16*): Radio and X-ray observations of SN1993J imply a mass-loss rate of $4 \times 10^{-5}$ $M_\odot$/yr and a wind-velocity of 10 km/s (*20*), close to an estimate of $2 \times 10^{-5}$ $M_\odot$/yr and wind velocity of 10 km/s consistent with the hydrodynamical state of the Cas A remnant (*16*). The presence of broad absorption components in the Cas A spectrum and the absence of prominent unresolved lines differ from typical spectra of SNe IIn whose ejecta are directly interacting with dense circumstellar gas. In the case of Cas A, the dense circumstellar wind may therefore not have extended directly to the surface and Cas A might have exploded into a bubble created by a temporary phase of enhanced wind velocity. In addition, evidence for CNO processing in the envelope of a red supergiant has been found both in the remnants of SN 1993J (*21*) and Cas A (*22*). A helium core mass of 3-6 $M_\odot$ inferred for SN 1993J (*12,13*) matches the total mass at core-collapse of Cas A based on the observational constraints for its remnant (*15*).

The optically bright emission of a SN IIb results from the production of a significant amount of 56-Ni, which for the SN 1993J supernova has been determined to be between 0.07 to 0.15 $M_\odot$ (*12, 13, 23*). Models of nucleosynthesis predict an associated amount of $(0.7 - 1.7) \times 10^{-4}$ M$_\odot$ 44-Ti produced for a stellar mass of 15-20 M$_\odot$ (*23*). A 44-Ti mass of $1.6^{+0.6}_{-0.3} \times 10^{-4}$ M$_\odot$ has been measured in the Cas A remnant (*24*). This is consistent with the Cas A supernova being an optically bright supernova such as SN 1993J.

Since the density and composition of the interstellar cloud giving rise to the echo is relatively unconstrained, it is difficult to accurately determine the peak brightness and light curve of Cas A from the scattered light. The circumstances regulating the infrared emission are simpler, and since the dust cooling time in the infrared echo is short (*5*), the rate of fading at 24 μm should be similar to the rate of fading of the heating energy pulse. The surface brightness at the peak position of the IR echo within 140 days between 20 Aug 2007 and 7 January 2008 faded by a factor of 18 +/- 3. This can be compared to a brightness decrease by a factor of 17 in the exponentially decaying bolometric light curve of SN 1993J (*25*) between day 33 and day 173.

It is a historical enigma whether Astronomer Royal Flamsteed witnessed the Cas A supernova on 16 August 1680 at sixth magnitude (*26*). For the peak visual brightness of -17.5 mag for SN 1993J (*25*) and a foreground extinction of $A_V \sim 8$ mag, a maximum visual brightness of 3.2 mag would be predicted. This value and the rapid decay (e.g., to sixth magnitude in only two months) are consistent both with the lack of widespread reportage, indicating the peak was fainter than third magnitude (*15*), and Flamsteed's observation. The visual extinction varies across the remnant and we have considered here the most likely extinction value within the plausible range at the center of Cas A (27).

One aspect about Cas A remains puzzling: The progenitor of SN 1993J was a binary star as now confirmed by the detection of a companion (*28*). Interestingly, a progenitor of 15-25 M$_\odot$ that loses its hydrogen envelope to a binary companion and undergoes an energetic explosion is more consistent with the theoretical models and observational constraints for Cas A than an evolved single star undergoing core collapse (*15*). Although there is at present no evidence for a companion that has survived the explosion (*27*), it might be speculated that two binary companions have merged during a common envelope phase before the explosion (*29*). The observed asymmetric distribution of the quasi-stationary flocculi (*18, 30*) near Cas A might originate from the loss of such an envelope.

Finally, we address one difference between the Cas A supernova spectrum compared to SN 1993J, the presence of two unresolved emission lines at 8727 and 9850 Å which we suggest are from neutral carbon in the interstellar echo cloud. The cloud is located at a distance of 266 +/- 23 light yr to Cas A [supporting online text]. For the light curve of SN 1993J (*25*), the peak visual brightness of the Cas A supernova at the location of the cloud is -12.9 mag, slightly brighter than Full Moon, and leading to a flux density of $F_{5600Å}$ ~4.9 × 10$^{-4}$ erg cm$^{-2}$ s$^{-1}$ A$^{-1}$ which is about 1000 times stronger than the interstellar radiation field in the solar vicinity. The color temperature of SN 1993J near maximum light was close to 10,000 K (*25*). While the relative flux of Lyman photons capable of ionizing hydrogen is low at such a color temperature, the excitation of carbon lines can still be significant (31): Carbon is by a factor of 10 the most abundant atom which can be ionized by photons less energetic than the Lyman limit. Emission lines of carbon are therefore ubiquitous in the predominantly neutral shielded environments of dense interstellar clouds irradiated with UV-optical radiation (31, 32). The [C I] transitions at 8727 and 9850 Å lines are known to be the brightest carbon emission lines of photon dominated regions in the optical wavelength range (32), however these lines have so far never been observed as the most prominent lines in SNe. Thus, it appears likely that the carbon lines are not intrinsic to the supernova spectrum but excited by the UV-optical supernova flash in the echo cloud.

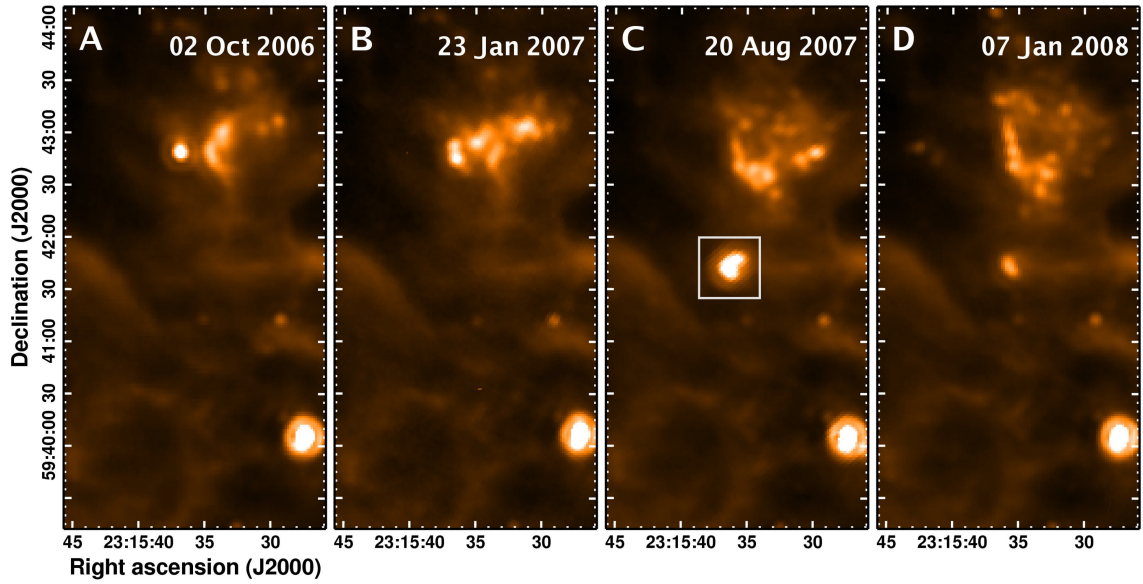

Fig. 1. Infrared images of the echo region. The panels A-D show MIPS 24 μm images of the same area of 2.5 x 5 arcmin² with the corresponding observing epoch labelled on each panel. The bright infrared echo is visible at the center of panel C. Other infrared echoes ~ 60 arcsec north of this feature are indicated by strong morphological changes in the time series, note e.g. the compact echo in panel A which disappeared later, in contrast to the smooth and unchanged interstellar cirrus emission. The white rectangle in panel C denotes the size of the optical images of the bright echo region shown in Fig. 2.

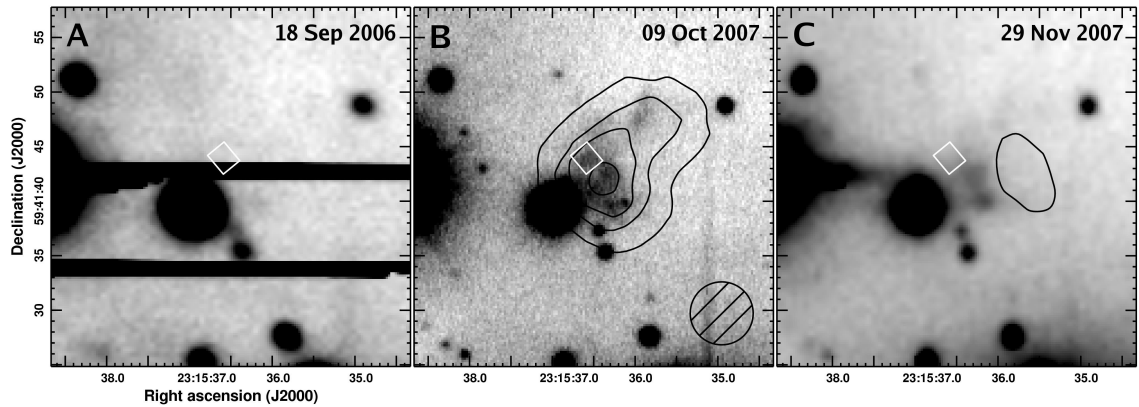

Fig. 2. Optical images of the echo region. Panels A-C show R-band images of the same area of 33 x 33 arcsec² with the corresponding observing epoch labelled on each panel. The white rectangle shown in all three panels denotes the 2 x 2.1 arcsec² extraction aperture of the Cas A supernova spectrum. Seeing was 1.6, 0.7 and 1.3 arcsec, FWHM, for panels A,B and C, respectively. The contours in panel B denote 24 micron emission from Fig. 1C in steps of 25 MJy/sr starting at 55 MJy/sr. The size of the MIPS 24 μm beam is indicated in the lower right. The contours in panel C display 24 μm emission from Fig. 1D with the same levels plotted in panel 2B.

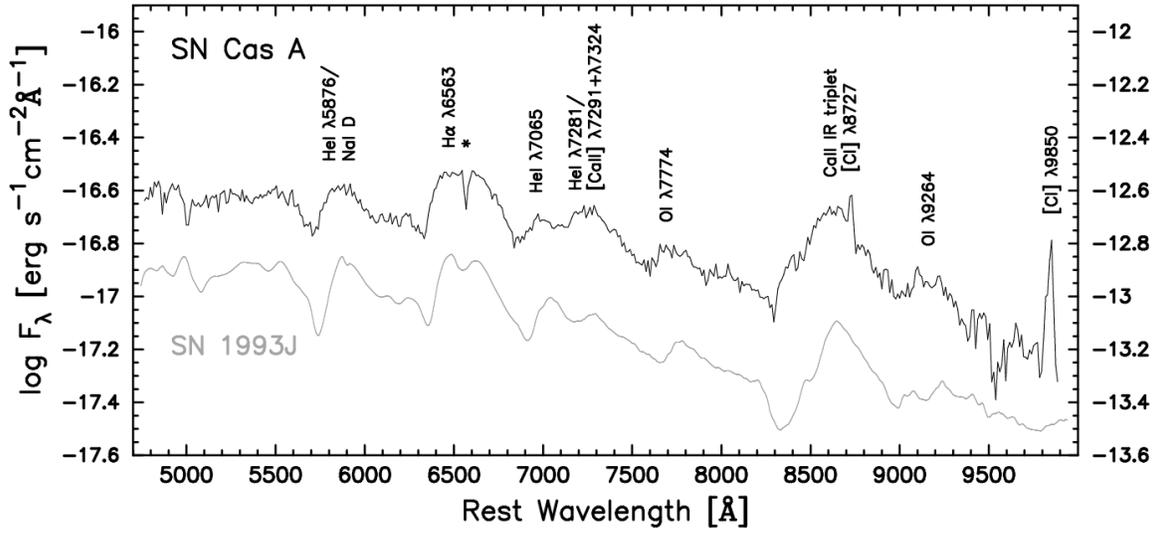

Fig. 3. Spectrum of the Cas A supernova and SN 1993J. Spectral features are labelled with their rest wavelength given in Å. The spectrum was extracted from the aperture shown in Fig. 2 and binned to 11.2 Å pixel$^{-1}$. Details of the data reduction are described in the supporting online text. The comparison spectrum of SN1993J was dereddened using E(B-V) = 0.2 mag (Ref. 28) and shifted by $\log(F_\lambda) = 4$.

## Supporting Online Text

**Spectroscopic observations**

Flux calibration was performed against the standard star G191B2B which was observed at comparable airmass. Atmospheric absorption features were removed using the stellar spectrum of a star of spectral type A0 observed in the same slit as the echo. The uncertainty of the flux calibration is 15%. The absolute flux calibration at wavelengths greater than 9100 Å and the removal of $H_2O$ absorption around 9500 Å are not optimum due to second order contamination of the standard star spectra and Paschen lines of the comparison star. The spectrum was dereddened to account for interstellar extinction, using an E(B-V) = 1.0 +/- 0.2 mag derived from the Balmer decrement of the underlying HII region Sharpless 157. The background spectrum of the HII region was subtracted. The sharp absorption feature (marked by an asterisk) on top of the broad H line is an artefact from the removal of this background spectrum. The color dependence of scattering was taken into account for a scattering angle $\theta$ = 104° (see Fig. S1) according to Draine (2003) and using the differential scattering cross sections provided by Draine (2008). The total integration time for the Cas A supernova spectrum was 5.5 h.

The comparison spectrum was derived as time-average of brightness-weighted spectra of SN 1993J obtained during days 1 to 83 after collapse. We have used spectra from the following epochs for this average (date of explosion, 1993 March 27 UT; JD 2,449,074): Days 2,3,7,9, from Clocchiatti et al. (1995) and days 19, 34, 45, 56 and 83 from Matheson et al. (2000). For days 2 and 3 we assumed blackbody spectra with photospheric temperatures of 55,000 and 30,000 K, respectively (Clocchiatti et al. 1995). The resulting spectrum is mostly dominated by the spectrum close to the strong secondary brightness peak (day ~ 20).

**Echo Geometry**

All echoes corresponding to light of an epoch $t$ after outburst are on an ellipse with Cas A and Earth at its foci (see Fig. S1). For any point of such an ellipse the distance (Cas A – echo) + (echo – Earth) = (Cas A – Earth) + c $(t + T_0)$, where $T_0$ is the time since light from the explosion first reached earth (327 +/- 19 yr), $D$ the Cas A distance ($3.4^{+0.3}_{-0.1}$ kpc) and $c$ the speed of light. The distance and scattering angle of the echo knot in Fig. 2B are d = 266 +/- 23 light yr and $\theta$ = 104 +/- 8°. The echo knot diameter on the sky is $\Phi$ = 2 arcsec, corresponding to 0.1 light yr. The time difference of 122 days between $t_0$ and 29 Nov 2007, when the echo had significantly faded, corresponds to 0.25 light yr along line of sight, indicating an elongated cloud.

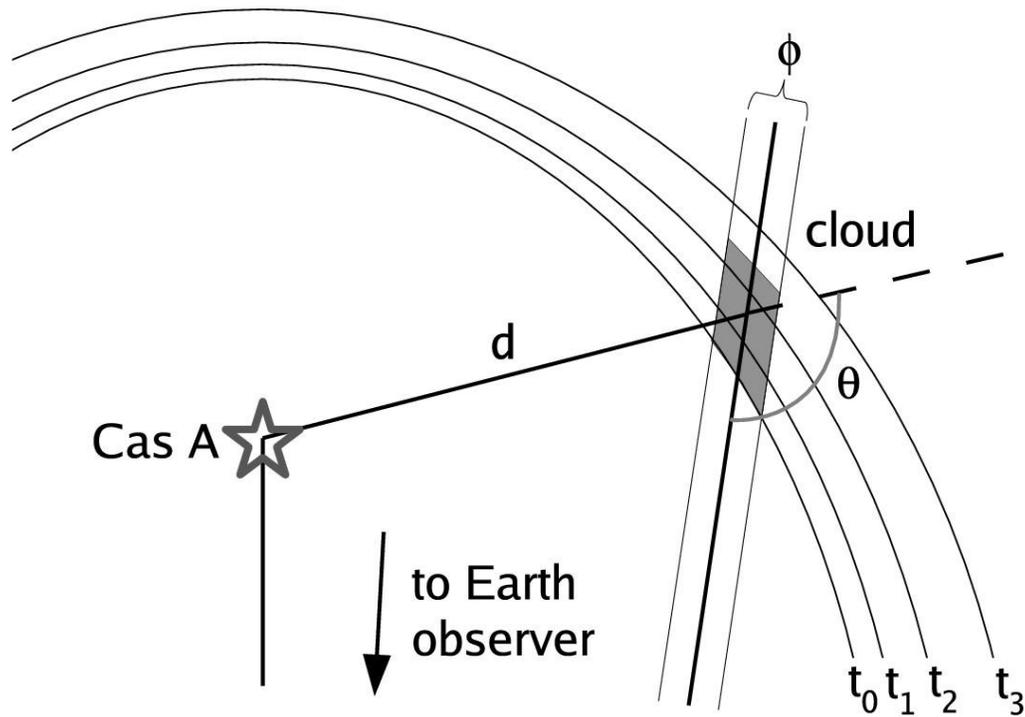

**Figure S1.** Sketch of the echo geometry. Ellipses corresponding to the leading edge of the supernova light pulse are shown for the following points in time: Encounter with the cloud boundary ($t_0$; unobserved, estimated 19 July 2007), *Spitzer* image on 20 August 2007 ($t_1$), optical spectrum on 9 Oct 2007 ($t_2$), and *Spitzer* observation on 7 Jan 2008 ($t_3$).

**Possibility of an unrelated galactic or extragalactic supernova**

The observed echo spectrum has a color of E(B-V) = 1.6 +- 0. The intrinsic color of SN IIb 1993J near maximum light is within the range 0.0 < E(B-V) < 1.0 (Richmond et al. 1994). Assuming the spectrum would originate from an unrelated supernova, the color of the spectrum would be the sum of intrinsic supernova color plus reddening due to extinction, and constraining the visual extinction to $A_V < 5.0 +- 0.3$ according to Rieke & Lebofsky (1985). From the typical SN IIb brightness of $M_V = -17.5$ (Richardson et al. 2006) and the observed integrated brightness of the echo feature $m_V = 23.3$, we derive a distance modulus of m-M > 35.8 and thus require a supernova distance d > 140 Mpc. If an extragalactic supernova at this distance would be the light source of the spectrum, all spectral features should exhibit a significant redshift which is not observed. On the other hand, a galactic supernova (with d < 40 kpc) would be either much brighter ($m_V < 5.5$ mag instead of 23.3 mag) or require a large extinction $A_V > 22.8$ which again is not consistent with the observation.

## Supporting References